\documentclass[a4paper,12pt]{article}
\usepackage[dvipdfmx]{graphicx}
\usepackage{cite}
\usepackage{color}

\title{Precision Measurement of the Position-space Wave Functions
of Gravitationally Bound Ultracold Neutrons}
\author{
Y. Kamiya\footnote{email: kamiya@icepp.s.u-tokyo.ac.jp},
G. Ichikawa\footnote{
currently at Department of Physics, Nagoya University}, and S. Komamiya\\\\
{\small {\it Department of Physics, Graduate School of Science,}} \\
{\small {\it and International Center for Elementary Particle Physics,}}\\
{\small {\it The University of Tokyo, Tokyo 113-0033, Japan}} }

\date{}
\begin{document}
\maketitle
\subsection*{Abstract}
Gravity is the most familiar force at our natural length scale.
However, it is still exotic from the view point of particle physics.
The first experimental study of quantum effects under gravity
was performed using a cold neutron beam in 1975.
Following this, an investigation of gravitationally bound quantum states
using ultracold neutrons was started in 2002.
This quantum bound system is now well understood,
and one can use it as a tunable tool to probe gravity.
In this paper, we review a recent measurement 
of position-space wave functions of such gravitationally bound states,
and discuss issues related to this analysis,
such as neutron loss models in a thin neutron guide, 
the formulation of phase space quantum mechanics,
and UCN position sensitive detectors.
The quantum modulation of neutron bound states
measured in this experiment
shows good agreement with the prediction from quantum mechanics.

\subsection*{Introduction}
Phenomena due to the gravitational field 
have been well understood at the macroscopic scales.
However, there are only a few cases of experiments at microscopic scales,
due to gravity's extreme weakness when compared to the other forces, 
such as electromagnetic and nuclear forces. 

Even though quantum mechanics was established in the early 1900s,
the first experiment to investigate a quantum effect under gravity 
was reported in 1975 by the group of R. Colella \cite{cow1},
in where a neutron interference pattern induced by a gravitational potential was observed.
The major systematic uncertainty was attributed to a bending effect of an interferometer 
when one rotates the system to manipulate a relative quantum phase difference
between two neutron paths.
A subsequent series of experiments were carried out 
with careful studies of systematic effects 
by measuring the bending effect using X-rays and considering the Sagnac Effect \cite{cow2}
due to the rotation of the Earth \cite{cow3,cow4}. 
The latest attempt was performed using normal-symmetric and skew-symmetric interferometers 
and a two-wavelength difference measurement method. 
The agreement with theoretical calculation is at the 1\% level \cite{cow5}.

Another approach to observe quantum effects under gravity 
was carried out by the group of V. V. Nesvizhevsky \cite{nesv1,nesv2}.
In this experiment, one measured a transmission of ultracold neutrons (UCNs)
through a vertically thin neutron guide in the terrestrial gravitational field
as a function of the slit thickness,
and showed the evidence of quantum bound states in the gravitational potential 
by analyzing the minimum height of a guide through which UCN can be transmitted.
In a later detailed analysis, the characteristic sizes of position-space wave functions 
for the first and second quantum states were evaluated with an uncertainty of around 10\% \cite{nesv3}.
The major systematic uncertainty in this evaluation came from the difficulty 
in modeling the rough surface of a scatterer used on the ceiling of the guide,
which causes some deformation of the wave functions.
To minimize the effects of the deformation, an ``differential method" was designed, 
in which one directly measures the wave functions using a position sensitive detector.
In this method, a neutron guide of moderate height can be used and
the deforming effect was kept rather small compared to the previous ``integrated method".
The first attempt of the differential method was reported in \cite{nesv3},
where a plastic nuclear track detector (CR39) 
with an uranium coating \cite{nesv4} was used as the detector.

\begin{figure}[htbp]
  \begin{center}
  \includegraphics[width=10cm]{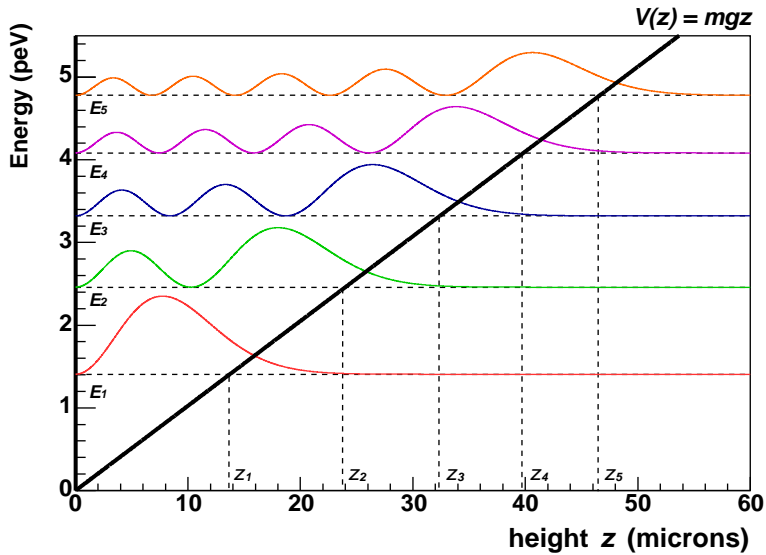}
  \end{center}
  \caption{Probability distributions of gravitationally bound states of neutrons
  for the first five states\cite{tokyo4}. 
  Horizontal lines indicate eigenenegies $E_n$ 
  and vertical lines show the classical turning points $z_n$ for each state.
  }
  \label{Airy}
\end{figure}

These quantum systems, in which the UCN bounces on a smooth floor, 
are expressed by the time-independent Schr\"{o}dinger equation,
\begin{equation}
\{ -\frac{\hbar^2}{2m}\frac{\mathrm{d}^2}{\mathrm{d} z^2} + V(z)\} \psi_n(z) = E_n \psi_n(z) \label{schrodinger}
\end{equation} 
with the linear potential
\begin{equation}
V(z) = \left\{
  \begin{array}{ll}
  m gz, & z \ge 0,\\
  \infty, & z \le 0,
  \end{array}
\right.
\end{equation}
where $\psi_n$ and $E_n$ are the eigenfunctions and eigenenergies,
$\hbar$ is the reduced Planck constant, $m$ is the neutron mass, 
and $g$ is the standard gravitational acceleration.
Equation (\ref{schrodinger}) can be rewritten in dimensionless form as
\begin{equation}
(\frac{\mathrm{d}^2}{\mathrm{d} \xi_n ^2} - \xi_n)\psi_n(\xi_n) = 0,
\end{equation}
where $\xi_n \equiv z/z_0 - E_n/E_0$
and the characteristic length and energy are given by
\begin{eqnarray}
z_0 & = & (\frac{\hbar^2}{2m^2g})^{1/3} ~~~~ \sim 6 ~ \mu\rm{m,} \label{nondimz}\\
E_0 & = & (\frac{m g^2 \hbar ^2}{2})^{1/3} ~~ \sim 0.6 ~ \rm{peV.} \label{nondimE}
\end{eqnarray}
This is the Airy equation and solutions are described by Airy special functions,
 $\rm{Ai}(\xi_n)$ and $\rm{Bi}(\xi_n)$.
Calculated probability distributions for the first five states are 
illustrated in Fig. \ref{Airy}.
Note that $\xi_n = 0$, where $z_n \equiv z_0 E_n/E_0$, 
are the classical turning points of a classical bouncing ball,
which are denoted in the cross points of each eigenenergy and the potential line of $mgz$ in Fig. \ref{Airy}.

This system is a suitable device with which to test the inverse square law of 
standard Newtonian gravity
and to search for new gravity-like short-range interactions around these scales.
The first limit to a non-Newtonian force with a Yukawa-type interaction potential 
using this microscopic system was reported in \cite{nesv5}.
The limits for a CP-violating Yukawa-type potential were shown in \cite{nesv6,nesv7}.
Several experimental schemes have been proposed to improve
the sensitivity to such hypothetical new physics.
One idea to achieve a better resolution on the measurement of 
the characteristic length is 
to utilize a convex reflection mirror to magnify the neutron distributions
\cite{nesv8,tokyo1}.
In this review, we discuss an experiment using this scheme to
precisely measure the UCN position-space wave functions,
performed by the group of S. Komamiya of the University of Tokyo\cite{tokyo2}.
Experiments exploiting other ideas for measuring the energy scale, 
i.e. the energy differences between quantum states, 
are proposed by observing resonance transitions 
induced by a magnetic field\cite{granit1,granit2,granit3,granit4} 
and mechanical vibrations\cite{qbounce3}.
These projects are called GRANIT and qBounce, respectively.
The first measurement of the resonance transition from the ground to the third state
was reported by the group of H. Abele in \cite{qbounce4}\footnote{
Recently, a new limit for the CP-Violating Yukawa-type potential using
the resonance method was reported in \cite{chameleon1}.
It also shows a limit for the chameleon field\cite{chameleon2,chameleon3,chameleon4}, 
a dark energy candidates.}.
For details of those projects, please see review papers
\cite{granit} and \cite{qbounce}
in this special issue.

\subsection*{Precision measurement and issues related to this analysis}
\begin{figure}[htbp]
  \begin{center}
  \includegraphics[width=10cm]{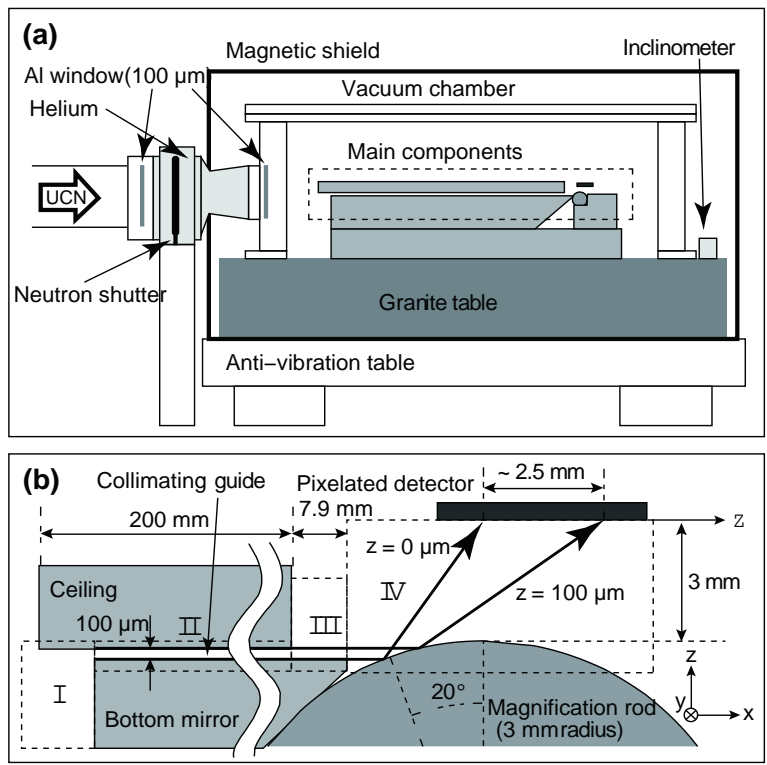}
  \end{center}
  \caption{Schematic drawing of the precision measurement experiment using
  a convex magnification mirror\cite{tokyo2}. 
  (a) is an overview of the system and (b) shows the geometry of the main components.
  }
  \label{setup}
\end{figure}

The precision measurement with a convex magnification mirror\cite{tokyo2} 
was performed using a UCN source provided at ILL(Institut Laue-Langevin)\cite{source1}.
The velocity distribution was measured by a TOF method and 
is well characterized by a Gaussian distribution with mean of 9.4 m/s and standard deviation of 2.8 m/s.
Figure \ref{setup}(a) shows a schematic drawing of the experiment.
The entire system was mounted on an anti-vibration table 
and magnetically shielded by a Permalloy sheet.
The main components can be separated into three parts: 
a vertically thin collimating guide, a magnification mirror(rod),
and a pixelated position sensitive detector.
The layout of these components is shown in Fig. \ref{setup}(b).

\subsubsection*{Collimating guide}
\begin{figure}[htbp]
  \begin{center}
    \includegraphics[width=10cm]{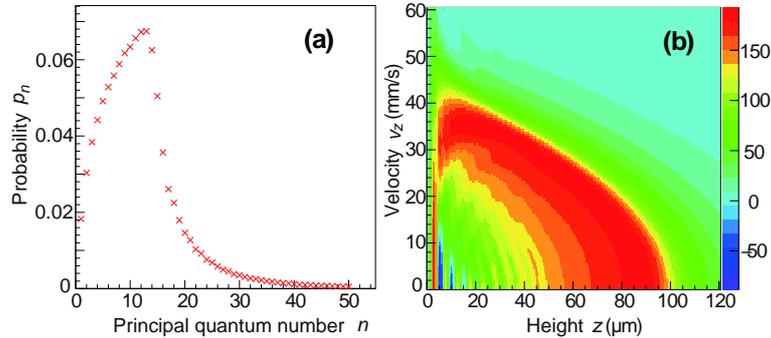}
  \end{center}
  \caption{(a) is an estimated population distribution (probability distribution of the guide transmission) 
  at the end of region I\hspace{-.1em}I\hspace{-.1em}I, and (b) shows the corresponding Wigner phase space distribution\cite{tokyo2}.
  }
  \label{wigner}
\end{figure}

The guide settles gravitationally bound quantum states.
To clearly distinguish each quantum state, which have energy differences of order of $\Delta E \sim 1 ~$peV,
a resolving time of $\Delta t \sim \hbar/\Delta E \sim 1$ msec is required.
Considering that the horizontal velocities of UCNs which we use are less than $15$ m/s,
the collimating guide was designed to have a sufficient length (192 mm in our setup)
to form each quantum states.
The bottom mirror is made of polished glass with a roughness of 
$R_a$(arithmetic average) $= 0.03 ~\mu$m.
The ceiling scatterer is a Gd-Ti-Zr alloy (54/35/11) deposited on glass, 
with a Fermi potential tuned to be nearly zero,
and a roughness of $R_a = 0.4 ~\mu$m\cite{tokyo3},
which scatters out neutrons in the higher states.
The guide selects lower states with appropriate populations 
to improve the contrast of the quantum spatial modulation\footnote{
The eigenfunctions modulate somewhat coherently,
especially around the lower region, except for the ground state, as seen in Fig. \ref{Airy}. 
To improve the contrast of the quantum modulation,
a method using a negative step of several tens of $\mu$m,
which transfers neutrons from the ground state to higher states,
was considered in \cite{nesv4}. Results were reported in \cite{nesv10,qbounce1,qbounce2}.}.
The neutron loss models for the scatterer or rough surface have been 
discussed in detail in \cite{nesv11,nesv12,nesv13},
and they are still interesting issues not only for UCN guiding applications but also for storage experiments 
such as neutron EDM measurements.
In the experiment reviewed in this paper, 
empirical models of the loss rates are adopted\cite{tokyo2}.
The loss rate by the scatterer, $\Gamma_n$, is assumed to be proportional 
to the probability of finding neutron in the roughness region,
and is given by
\begin{equation}
\Gamma_n = \gamma \int_{h-2\delta}^{h} | \tilde{\psi_n} |^2 dz ~,
\end{equation}
where $\gamma$ is a scaling constant,
estimated from data to be $9.5^{+0.7}_{-0.9} \times 10^4$ s$^{-1}$,
$h$ is the height of the guide ($100 ~ \mu$m),
$\delta$ is the roughness of the scatterer ($0.4 ~ \mu$m),
and $\tilde{\psi_n}$ are deformed wave functions in the guide.
Neutron losses at the bottom mirror
due to absorption, non-specular reflection, up scattering and other processes
are modeled empirically as
\begin{equation}
B_n = \beta \frac{g}{2\sqrt{2}}\sqrt{\frac{m}{\tilde{E_n}}} ~,
\end{equation}
in which the loss rate is assumed to be proportional to the classical bouncing number per unit time,
$g/2\tilde{v}_{z,n,max}$,
where $\beta$ is a scaling constant (estimated to be $0.38^{+0.04}_{-0.03}$), 
$\tilde{v}_{z,n,max} \equiv \sqrt{2\tilde{E_n}/m}$ is the maximum vertical velocity of a neutron in the $n$th state,
and $\tilde{E_n}$ is the eigenenergy of the deformed wave functions.
The transmissivity of the guide for each state can be
written as 
\begin{equation}
\tilde{p_n} \propto \left< \exp[-\frac{l}{v_x} (\Gamma_n + B_n)] \right>_{v_x} ~,
\end{equation}
where $<...>_{v_x}$ indicate the average over the neutron horizontal velocities.
By applying a diabatic transition from region I\hspace{-.1em}I to I\hspace{-.1em}I\hspace{-.1em}I,
the population distribution (a probability in the paper\cite{tokyo2}) 
of neutrons for each state at the end of region I\hspace{-.1em}I\hspace{-.1em}I is estimated as in
Fig. \ref{wigner}(a).

\subsubsection*{Magnification mirror}
\begin{figure}[htbp]
  \begin{center}
  \includegraphics[width=10cm]{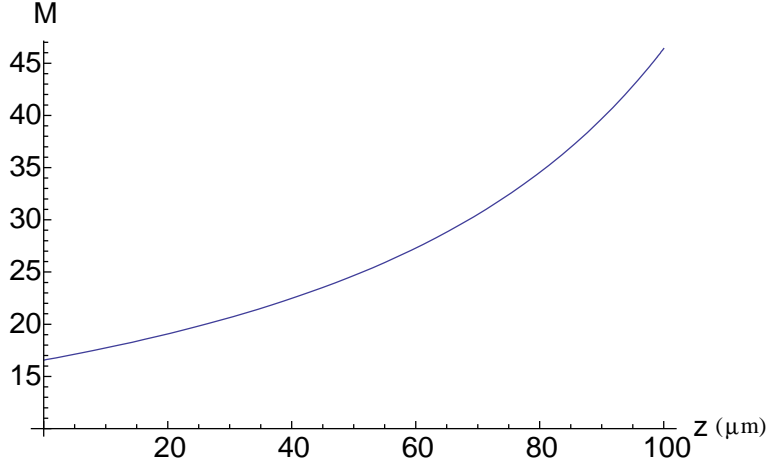}
  \end{center}
  \caption{Magnification power $M$ of the cylindrical rod as a function of the height $z$\cite{tokyo4}.
  It is about 20 times magnification around the lower region.}
  \label{mag}
\end{figure}

A Ni coated cylindrical rod is used as a magnifying convex mirror.
Its radius is 3 mm and it is placed to have a grazing angle of 20 deg. for
horizontally moving neutrons at the bottom floor level ($z = 0$).
Figure \ref{mag} shows the magnification power 
as a function of the height \cite{tokyo4}. 
It gives about 20 times magnification around the lower region of $z \sim 20 ~ \mu$m.
Before depositing the Ni coating, the cylinder was polished
at the Research Center for Ultra-Precision Science and Technology, Osaka University.
The roughness of the rod after the depositing was measured to be
$R_a = 1.9$ nm, two orders of magnitude smaller than the neutron wavelength.

\begin{figure}[htbp]
  \begin{center}
  \includegraphics[width=10cm]{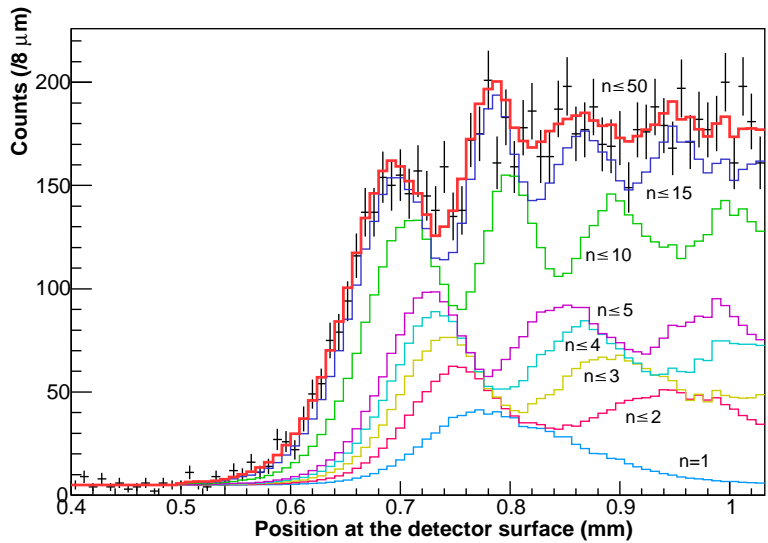}
  \end{center}
  \caption{Black crosses show measured neutron position distribution\cite{tokyo2}.
  Modulated distribution was clearly measured.
  The result shows good agreement with the quantum expectation 
  calculated using phase space formalization by Wigner distribution (p-value is 0.715). 
  The best estimated line is shown by a red solid curve and
  the other lines indicate contributions of each state to the distribution.
  For definitions of fitting parameters, best fit values, systematic uncertainties, 
  and the other details, see \cite{tokyo2}. }
  \label{result}
\end{figure}

To calculate the detected position distribution on the detector surface,
the wave function at the end of the bottom floor
is re-expressed by the Wigner phase space distribution\cite{wigner1,wigner2},
\begin{equation}
W(z,p_z) \equiv \frac{1}{2\pi\hbar} \int_{-\infty}^{\infty} \delta \eta ~ e^{-\frac{i}{\hbar} p_z \eta}
<z+\frac{1}{2}\eta | \hat{\rho} | z-\frac{1}{2}\eta> ~,
\end{equation}
where $p_z$ is the momentum and $\hat{\rho}$ is a density operator.
The Wigner distribution is known as a phase space formulation of quantum mechanics,
and is widely used for quantum optics, for example in the study of decoherence\cite{wigner3}.
As an application to massive particles, one can find a paper which shows 
a phase-space tomography of the Wigner distribution for a coherent atomic beam
in a double-slit experiment\cite{wigner4}.
Figure \ref{wigner}(b) shows the Wigner distribution 
constructed from the estimated populations in our experiment\cite{tokyo2}.
The time evolution of the Wigner distribution is calculated
by the evolution of the density operator described by the Liouville - von Neumann equation
\begin{equation}
\frac{\partial \hat{\rho}}{\partial t} = - \frac{i}{\hbar}[ \hat{H},\hat{\rho}] ~,
\end{equation}
where 
\begin{equation}
\hat{H} = \frac{\hat{p_z^2}}{2m} + V(\hat{z}) ~.
\end{equation}
By evaluating the kinetic part as
\begin{equation}
T = - \frac{p_z}{m}\frac{\partial}{\partial z} W(z, p_z) ~,
\end{equation}
and the potential part as
\begin{equation}
\sum_{l=0}^{\infty} U_l = \sum^{\infty}_{l=0}\frac{(-1)^l (\hbar/2)^{2l}}{(2l+1)!}
\frac{d^{2l+1}V(z)}{dz^{2l+1}}\frac{\partial^{2l+1}}{\partial p_z^{2l+1}} W(z,p_z) ~,
\end{equation}
one can obtain the Quantum Liouville Equation for the Wigner distribution \cite{wigner5}
\begin{equation}
\left( \frac{\partial}{\partial t} + \frac{p_z}{m}\frac{\partial}{\partial z} - \frac{dV(z)}{dz}\frac{\partial}{\partial p_z}
\right) W(z,p_z) = \sum^{\infty}_{l=1} U_l ~.
\end{equation}
In the case of $V(z) = mgz$, the right hand side of the equation vanishes and it becomes
the classical Liouville equation. 
Therefore, in region I\hspace{-.1em}V, 
one can treat the evolution of each phase point of the Wigner distribution
as a classical path under gravity. 
Figure \ref{result} shows the measured data and the best theoretical estimation 
using this model.
The corresponding p-value is 0.715, and the experimental data support
the quantum features described by the phase space formulation
using the Wigner distribution\cite{tokyo2}.

\subsubsection*{Position sensitive detector for UCNs}
\begin{figure}[htbp]
  \begin{center}
  \includegraphics[width=10cm]{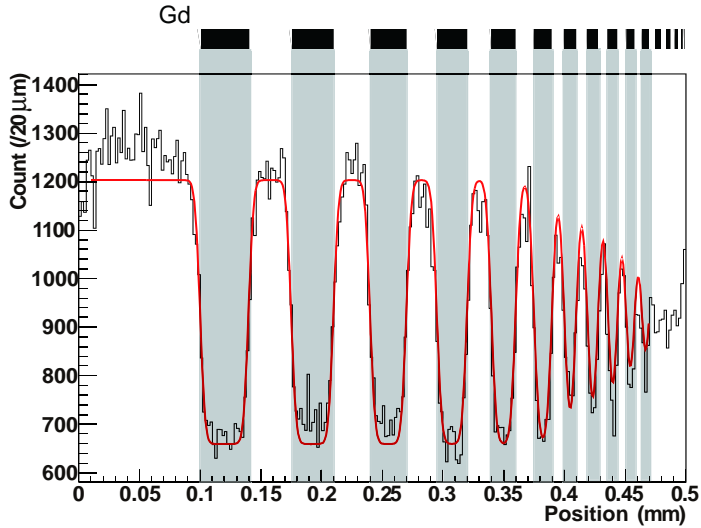}
  \end{center}
  \caption{Fitting results of a Gd shadow pattern\cite{tokyo4}.
  Spatial resolution is measured to be $3.35  \pm 0.09 ~\mu$m
  by evaluating these line spread functions. For details of evaluation method,
  see \cite{tokyo1,tokyo5}.}
  \label{ccd2}
\end{figure}

A back-thinned CCD (HAMAMATSU S7030-1008)
with thin Ti-$^{\rm{10}}$B-Ti layers
is used for the position sensitive detector in the experiment\cite{tokyo2}.
Its pixel size is $24~\mu$m $\times$ $24~\mu$m
and the thickness of the active volume is about $20~\mu$m.
The $^{10}$B layer converts neutrons into charged particles
by the nuclear reaction $^{10}$B(n,$\alpha$)$^{7}$Li.
The secondary particles are emitted in a nearly back-to-back configuration.
One of them deposits its kinetic energy in the active area
and creates a charge cluster, which typically spreads into nine pixels.
The weighted center of the charge cluster is a good estimation of the incident neutron position.
The thickness of the layers are 20 nm and 200 nm for Ti and $^{10}$B, respectively,
and they are formed by evaporating directly on the CCD surface.
The spatial resolution is measured to be $3.35 \pm 0.09 ~\mu$m
by evaluating the line spread function (LSF) (see Fig. \ref{ccd2}) 
using very cold neutron beams at ILL\cite{tokyo4}.
For the detail of the evaluation scheme, see \cite{tokyo1,tokyo5}.
A neutron converter of $^{6}$Li using $^{6}$Li(n,$\alpha$)$^{3}$H reaction was also investigated. 
It is concluded that the use of $^{10}$B gives better spatial resolution\cite{tokyo5}.

Other detectors using a silicon pixel device, Timepix\cite{timepix1}, 
with $^{6}$LiF and $^{10}$B converter were studied in \cite{timepix2}.
Its pixel pitch is $55~\mu$m $\times$ $55~\mu$m
and the thickness of its silicon layer is about $300~\mu$m.
The spatial resolution for the $^{6}$LiF converter was evaluated by LSF to be $2.3 ~\mu$m,
corresponding to about $5.3 ~\mu$m in FWHM of the point spread function (PSF).
The performance of the $^{10}$B converter is estimated 
by Monte-Carlo simulations to be better than $3 ~\mu$m in FWHM of the PSF.

Another detector concept of Uranium coated plastic nuclear tracker (CR39) \cite{nesv4} was used
in several experiments\cite{nesv3, nesv10, qbounce1, qbounce2, qbounce5}.
Usually two fission fragments are emitted from a thin $^{235}$U coating, 
and one of the daughter nuclei makes a track of defects in the CR39.
By chemical etching, the diameter of track points is increased up to 1 $\mu$m,
allowing us to scan these vertexes using an optical microscope.
Position resolution is around $1 ~\mu$m\cite{nesv4}.
By carefully analyzing the vertex shape, 
the spatial resolution can be improved to  $0.7 ~\mu$m\cite{qbounce5}.

\subsection*{Summary}
The quantum system of a gravitationally bound neutron
is one of the most suitable tools to investigate gravity
or gravity-like hypothetical interactions 
around the scale of $10 ~\mu$m in length or 1 peV in energy.
After establishing this research field 
by the pioneering work in observation of the quantum state\cite{nesv1},
experimental schemes for precision measurements of 
these characteristic scales have developed rapidly, 
and nowadays, one can establish limits for parameter spaces 
of new physics experimentally\cite{nesv5,nesv6,nesv7,chameleon1}.
Furthermore one can start discussing a phase space formulation of quantum physics 
for the gravitationally bound quantum state \cite{tokyo2}.
An experiment for a possible test of phase space formalization 
using spatial interference is under preparation.

\subsection*{Acknowledgments}
The authors would like to thank Valery V. Nesvizhevsky (Institut Laue-Langevin),
Hartmut Abele (Vienna University of Technology), William Snow (Indiana University),
Peter Geltenbolt (Institut Laue-Langevin), and all participants in the GRANIT-2014 Workshop
for interesting discussions and helpful suggestions. 
This material is based upon work supported by JSPS KAKENHI Grants No. 20340050 
and No. 24340045, and Grant-in-Aid for JSPS Fellows No. 22.1661.

\end{document}